\documentclass[usenatbib]{mn2e}
\usepackage{graphicx}

\def\gax{\mathrel{\raise.3ex\hbox{$>$}\mkern-14mu\lower0.6ex\hbox{$\sim$}}}
\def\lax{\mathrel{\raise.3ex\hbox{$<$}\mkern-14mu\lower0.6ex\hbox{$\sim$}}}
\def\gtorder{\mathrel{\raise.3ex\hbox{$>$}\mkern-14mu
             \lower0.6ex\hbox{$\sim$}}}
\def\ltorder{\mathrel{\raise.3ex\hbox{$<$}\mkern-14mu
             \lower0.6ex\hbox{$\sim$}}}

\begin{document}

\title[Dust Formation By Failed Supernovae]{Dust Formation By Failed Supernovae}

\author[C.~S. Kochanek]{ C.~S. Kochanek$^{1,2}$  \\
  $^{1}$ Department of Astronomy, The Ohio State University, 140 West 18th Avenue, Columbus OH 43210 \\
  $^{2}$ Center for Cosmology and AstroParticle Physics, The Ohio State University, 
    191 W. Woodruff Avenue, Columbus OH 43210
   }

\maketitle

\begin{abstract}
We consider dust formation during the ejection of the hydrogen envelope of a 
red supergiant during a failed supernova (SN) creating a black hole.  While
the dense, slow moving ejecta are very efficient at forming dust, only
the very last phases of the predicted visual transient will be obscured.  
The net grain production consists of $M_d \sim 10^{-2} M_\odot$ of very
large grains ($10$ to $1000\mu$m).  This means that failed SNe could be
the source of the very large extrasolar dust grains identified by {\it Ulysses},
{\it Galileo} and radar studies of meteoroid re-entry trails
 rather than their coming from an
ejection process associated with protoplanetary or other disks.  
\end{abstract}

\begin{keywords}
stars: supernovae:general: dust: black holes
\end{keywords}

\section{Introduction}
\label{sec:introduction}

The formation of black holes during a failed supernova (SN) has generally been
assumed to lead to a black hole with the mass of the star at the time core
collapse is initiated (e.g. \citealt{Heger2003}). \cite{Nadezhin1980} argued that this was not
true of red supergiants because the hydrogen envelope is so weakly bound
that the drop in the gravitational potential created by neutrino losses
is enough to unbind the envelope.  Radiation hydrodynamic simulations by
\cite{Lovegrove2013} confirmed this supposition.  Failed SNe of red 
supergiants lead to the formation of black holes with (roughly) the mass
of the helium core of the star at the time of collapse and not the total
mass of the star.  

This new expectation is of particular interest because studies of SN
progenitors appear to be finding too few massive progenitors (\citealt{Kochanek2008}) 
particularly in the upper end of the red supergiant mass range (\citealt{Smartt2009}).
\cite{Smartt2009} estimated that red supergiant progenitors were missing
from $(16.5\pm1.5)M_\odot$ to the upper mass limit for stars to explode
as red supergiants and become Type~IIP SNe ($25$-$30M_\odot$).  This is an interesting mass range
for failed SNe because theoretical studies indicate that many stellar models in
this mass range have density structures that make it more difficult to
produce a successful explosion (e.g. \citealt{Oconnor2011}, \citealt{Ugliano2012}).
Other options include changing stellar models so that stars in this mass
range do not end their lives as red supergiants (e.g. \citealt{Groh2013})
or using dusty winds to make them less observable (e.g. \citealt{Walmswell2012},
but see \citealt{Kochanek2012}).

Producing failed SNe in this mass range also provides a natural explanation for the observed
distribution of remnant masses (\citealt{Kochanek2013}).  The remnant
mass function is bimodal between neutron stars with masses near $1.4M_\odot$
and black holes with masses of $5$-$10M_\odot$, although it is uncertain
if the gap in the mass function is truly empty or simply a deep minimum
(e.g. \citealt{Bailyn1998}, \citealt{Ozel2010}, \citealt{Farr2011}, \citealt{Kreidberg2012}).
This gap is difficult to explain if black holes form by fall back of
material in a successful explosion (e.g. \citealt{Zhang2008}, \citealt{Fryer2012}). 
The gap is, however, a natural consequence of
combining supernovae without fall back, which best explain the masses
observed in neutron star binaries (\citealt{Pejcha2012}), with a population 
of $\sim 20M_\odot$ failed supernovae where the typical black hole mass of
$\sim 7M_\odot$ is simply the typical helium core mass of the
failed SNe.

\cite{Lovegrove2013} also pointed out that the ejection of the envelope
would be associated with a relatively long lived ($\sim$ year), modest
luminosity ($\sim 10^6 L_\odot$) transient largely powered by the 
recombination of the envelope.  We have been carrying out a search
for failed SNe using the Large Binocular Telescope following the 
ideas of \cite{Kochanek2008}, where we monitor nearby galaxies to
see if any stars effectively vanish, potentially with some intervening
transient phenomenon.  This survey can easily identify the transients
predicted by \cite{Lovegrove2013}, although it is not well suited for
identifying the shorter ($\sim$~week) shock break out peak associated
with the event (see \citealt{Piro2013}).

These more massive red supergiants tend to have winds that produce
silicate dusts and other products of oxygen-rich chemistry 
(see, e.g., the review by \citealt{Cherchneff2013}), and, as 
mentioned earlier, \cite{Walmswell2012} even propose using
this obscuration as a possible means of solving the red supergiant problem 
identified by \cite{Smartt2009}.  During
the envelope ejection following a failed SNe, this same material is
going to be ejected {\it en masse} at much higher densities and 
only moderately higher velocities, so presumably it is also likely to 
form dust but in much greater quantities.  This leads to the 
possibility that the optical properties of the transient are 
greatly modified from the dust free simulations carried out by
\cite{Lovegrove2013}.  

We already know of one class of transients that successfully cloaks themselves
in dust following their peaks, the SN~2008S class of explosive transients
from AGB stars (see, e.g., \citealt{Thompson2009}, \citealt{Kochanek2011}, \citealt{Szczygiel2012}).  
The mechanism here is very different, since the
dust in the SN~2008S class is almost certainly dust re-forming in a pre-existing 
dense wind after being destroyed by the transient (\citealt{Kochanek2011}).  However, it does
suggest an investigation of whether the visual transient predicted by 
\cite{Lovegrove2013} will be substantially modified by dust formation
and if the ejected dust has any consequences for our understanding of
the interstellar medium.
In \S2 we model dust formation in these systems to find that the
transients are not substantially modified even though they should
be very efficient at forming significant masses of very large dust
grains.  We discuss the implications in \S3.  

\section{Dust Formation }

A striking property of these transients is the ejection of the hydrogen 
envelope under conditions that seem remarkably conducive to the formation and
growth of dust grains.  The densities are very high compared to stellar
winds that easily form dust and the transient luminosities and temperatures
are relatively low.  Following our previous examination of dust formation
in Luminous Blue Variable (LBV) eruptions (\citealt{Kochanek2011b}), we adopt the physically reasonable view
that particle nucleation simply occurs once conditions allow 
nucleation, and that we must then follow the collisional growth of the grains
to estimate the resulting opacities and optical depths.
From \cite{Lovegrove2013} we have the transient temperature and luminosity,
$T_*(t)$ and $L_*(t)$, and the density and velocity profile at some late
time $t_0$ for their $15M_\odot$ models with ejecta kinetic energies
of $3.8$ and $8.9 \times 10^{47}$~ergs.  
At these late phases, we can view the ejecta as being in
free expansion, so the velocity $v_p(r_p)$ is now time independent.
Hence, the radius of the material at some other time $t$ is simply
$r(t) = r_p + (t-t_p) v_p(r_p)$ and at late times the density $\rho(t)$ is related
to the mass of the zone by 
$ m = 4 \pi v_p^2 \Delta v_p t^3 \rho(t) = 4 \pi (\Delta v_p/v_p) r_p(t)^3 \rho(t) $
where $m$ is the mass in the zone and $\Delta v_p$ is the velocity
difference across the zone edges.  

\begin{figure*}
\includegraphics[width=4.5in]{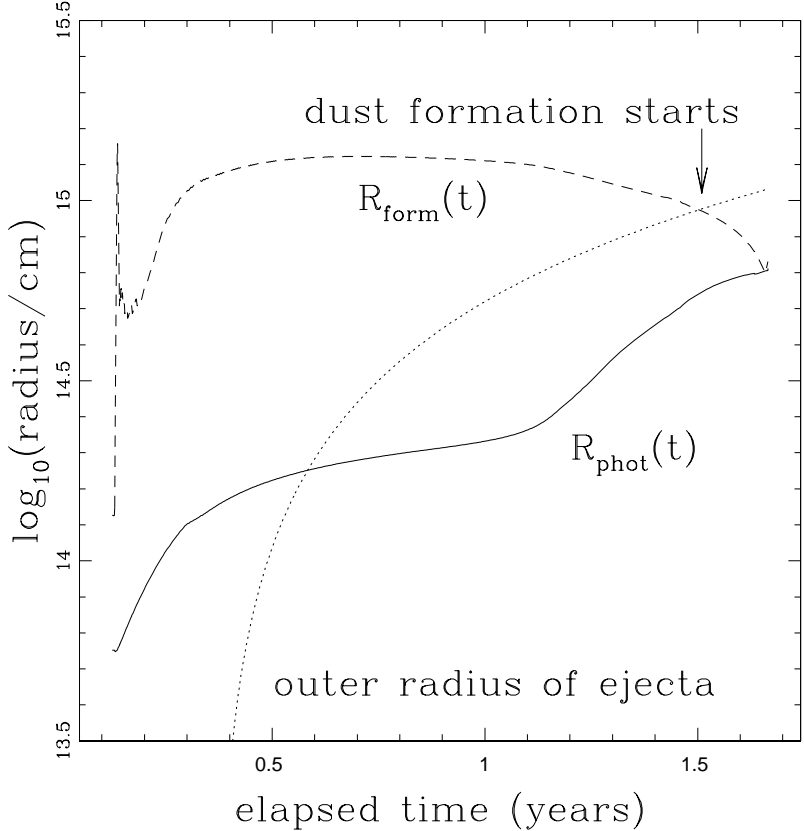}
\caption{
  Evolution of the photospheric radius $R_{phot}(t)$ (solid), the outer
  radius of the ejecta (dotted) and the dust formation radius 
  $R_f(t)$ (dashed) for the lower energy transient.  Dust formation 
  starts after 1.5 years when the outer radius of the ejecta
  becomes larger than $R_f$, as indicated by the arrow. 
  Here we used $T_f=1200$~K and silicate dust.  The outer
  radius simply uses the final velocity without the 
  accelerations that would make it converge with 
  $R_{phot}$ at early times.
  }
\label{fig:dform}
\end{figure*}

\begin{figure*}
\includegraphics[width=4.5in]{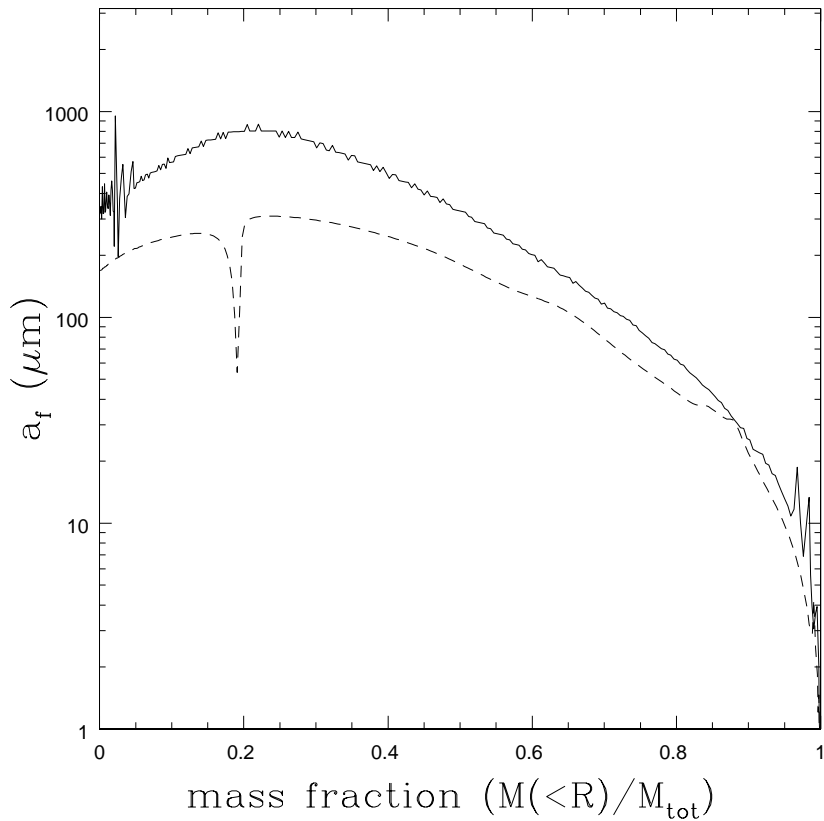}
\caption{
  Final silicate grain sizes $a_f$ as a function of the ejecta mass
  fraction $M(<R)/M_{tot}$ for the low (solid) and high (dashed)
  energy transients.  The fine structures in the curves are due
  to structures in the velocity difference $\Delta v$ across
  the Lagrangian zones of the \protect\cite{Lovegrove2013} simulations.
  Here we used $T_f=1200$~K.
  }
\label{fig:size}
\end{figure*}

\begin{figure*}
\includegraphics[width=4.5in]{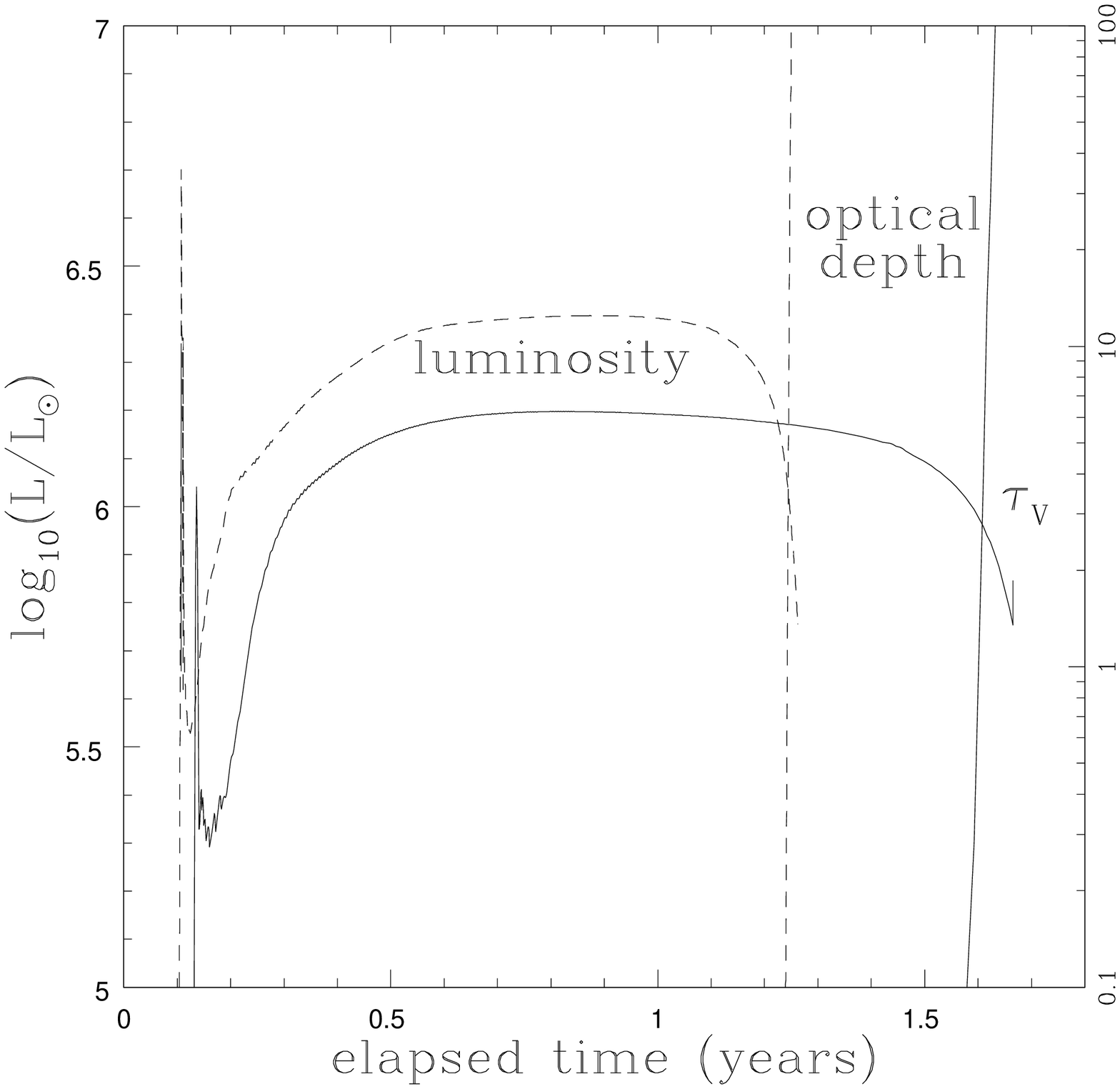}
\caption{
  The transient luminosity from \protect\cite{Lovegrove2013} (right scale)
  and the dust visual optical depth $\tau_V$ (right scale) for the
  low (solid) and high (dashed) energy transients.  As soon as 
  dust formation becomes feasible, the optical depth becomes
  ``infinite'' and will black out the already rapidly fading
  transient.  The initial luminosity spike occurs 
  when the low velocity shock wave causing the ejection
  first reaches the stellar surface (see \protect\citealt{Piro2013}).
  Here we used $T_f=1200$~K and silicate dust
  }
\label{fig:tau}
\end{figure*}

Since the radiation from these transients is cold ($T_* \sim 3000$-$4000$~K, \citealt{Lovegrove2013}), there
is no ultraviolet radiation that can stochastically heat dust grains as they
try to form (this is the mechanism that prevents dust formation around hot stars,
see \citealt{Kochanek2011b}).  We need only consider
whether the equilibrium temperature of a small grain is lower than the
temperature at which grains can form.  This
leads to the requirement that dust can only form once the ejecta
is more distant than the time-dependent dust formation radius 
\begin{equation}
     R_f(t) = \left( {  L_*(t) Q_P(T_*(t)) \over 16 \pi Q_P(T_f) \sigma T_f^4 }\right)^{1/2}
    \label{eqn:rform}
\end{equation}
where $Q_P(T)$ is the Planck mean absorption factor and $T_f\sim 1000$
to $2000$~K is the dust formation temperature. We use the 
values for the smallest 
grain sizes ($a=10^{-3}\mu$m) given by \cite{Draine1984} and \cite{Laor1993}.
Figure~\ref{fig:dform} compares the evolution of $R_f(t)$ to the photosphere
defined by $R_{phot}(t) = (L_*(t)/4\pi \sigma T_*(t)^4)^{1/2}$ and
to the radius of the outermost ejecta.   We find that the ejecta
are inside the $R_f(t)$ until very late in the transient, so the
first key result is that dust formation can only affect the late
phases of the transient.

Note that if the gas is cooling adiabatically, the gas temperature is 
essentially guaranteed to be lower than the dust formation temperature 
at the time dust formation commences.  We can normalize the gas temperature
such that it has the photospheric temperature when it is at the photosphere,
so $T_{gas}(R) = T_* (R_{phot}/R)^2$ (for $\gamma=5/3$).  When the gas
temperature reaches the formation temperature, $T_{gas}=T_f$, then the
dust temperature (ignoring Planck factors) due to the radiation field
of  $T_d=T_f (T_*/T_f)^{3/4} > T_f$ will always exceed the dust formation
temperature unless there is a precipitate drop in the luminosity.  We
explicitly checked this point as part of the calculation including the
time variability of the luminosity and found that dust formation was always
limited by radiative heating rather than adiabatic cooling.

We simply assume that nucleation occurs, and then allow
the dust to grow collisionally. The grain radius grows as
\begin{equation}
   { d a \over d t} = { v_c X_d \rho \over 4 \rho_{bulk} }
   \label{eqn:rate}
\end{equation}
where $v_c$ is the typical collisional velocity, $X_d$ is the
mass fraction of the species condensing onto the grains, $\rho$
is the gas density, and $\rho_{bulk}\simeq 2.2$ or $3.8$~g~cm$^{-3}$
is the bulk density of the graphitic or silicate grains (e.g.
\citealt{Kwok1975}, \citealt{Deguchi1980}).
If we normalize the density to the value at the formation radius 
$\rho = \rho_f(t_f/t)^3$ and scale the collision
velocities as $v_c = v_{cf}(t_f/t)^n$, then we can integrate
Equation~\ref{eqn:rate} to get the final size of the grains
\begin{equation}
    a_f = { v_{cf} R_f X \rho_f \over 4 (n+2) v \rho_{bulk} }.
     \label{eqn:afinal}
\end{equation}
The size at any intermediate time is simply
\begin{equation}
   a(t) = a_f \left[ 1 - \left( { R_f(t_f) \over r(t) } \right)^{n+2}\right]
\end{equation}
where $t>t_f$ and $t_f$ is the time the fluid element started to form dust.
Because growth is a collisional process, the rapidly dropping
density means that most of the growth occurs between $R_f$ 
and $2R_f$.  As a result, the choice for the velocity scaling
exponent $n$ is unimportant given the other uncertainties and we will
simply use $n=0$.  Here we have used a sticking probability of unity
and ignored coagulation, where the former slows and the latter speeds growth.
We will scale the amount of condensible
material to $X = 10^{-3} X_3$, since silicate dust formation
is more likely than carbonaceous dust given the properties
of the winds of these stars.  Given the temperatures near the
dust formation radius, we adopt $v_c = 1$~km/s as a reasonable
collision velocity. Changing these values allows moderate
changes in the grain sizes but leads to no qualitative changes.

As a check, we can apply this approach to a wind with mass loss rate 
$\dot{M}=10^{-5}\dot{M}_5M_\odot$~year$^{-1}$ and velocity $v_w=10 v_{w10}$~km/s 
around a star with luminosity $L_*=10^4L_{*4} L_\odot$.  A
silicate grain has an equilibrium temperature $T_f=1200$~K 
at $R_f \simeq R_{f14} 10^{14}$~cm, implying a final grain size of
\begin{equation}
    a_f={ \alpha v_c X \dot{M} \over 16 \pi \rho_b v_w^2 R_f }
  \simeq 0.01 { \alpha \dot{M}_5 X_3 v_{c1} \over R_{f14} v_{w10}^2 } \mu\hbox{m}
  \label{eqn:agb}
\end{equation}  
for a constant velocity wind with $\rho=\dot{M}/4\pi v_w r^2$.
This is the correct order of magnitude for dust formation around
an asymptotic giant branch (AGB) star with these properties.  For
example, \cite{Nanni2013} find $a_f \simeq 0.1\mu$m for $X_3\simeq 3$,
while our simple expression would give $a_f \simeq 0.03\mu$m. The
estimate is low because the constant velocity assumption
underestimates the time spent near $R_f$ as the formation
of dust accelerates the flow.  For dust formation in the
ejecta of a transient, this issue does not arise because
the velocities are not controlled by the dust formation process.

Figure~\ref{fig:size} shows the final grain sizes for silicate
dust formed by failed SNe assuming $T_f=1200$~K.  
Graphitic grains are larger roughly
by the ratio of the bulk densities.  The smaller scale 
structures in Figure~\ref{fig:size} are caused by fluctuations
in the velocity divergence $\Delta v$ of the individual
Lagrangian zones in the velocity profile from \cite{Lovegrove2013}
that would average out over the expansion time but are 
present in an instantaneous snap shot of the ejecta.
As expected, these transients offer excellent conditions for
the growth of grains, with peak sizes for our nominal
parameters on the scale of millimeters.  With $M_{tot}=8M_\odot$
of ejecta, the net dust production is 
$M_{dust} = X M_{tot} \sim 10^{-2}M_\odot$ of dust.
There is some freedom to adjust the final sizes by 
changing the parameters ($X$, $v_{cf}$, $n$) in 
Equation~\ref{eqn:afinal}, but not enough freedom to
change the qualitative result that failed SNe should produce
a significant mass of very large dust grains.

At any given time, we can now determine which of the Lagrangian 
zones from the \cite{Lovegrove2013} simulations have passed their
dust formation radius, and the size to which the grains have 
grown since that time.  We then calculate the visual (V-band) 
opacities of these layers using the scaled cross sections $Q_V(a)$. 
The optical depth is then
\begin{equation}
     \tau(t) = \int dr { 3 X \rho(r,t) Q(a(r,t)) \over 4 \rho_{bulk} a(r,t)} 
\end{equation}
where the integral extends from the dust formation radius $R_f(t)$
from Equation~\ref{eqn:rform} to the
outer edge of the star, and the density $\rho(r,t)$ and grain size $a(r,t)$
are functions of both radius and time.  We separately calculate
the absorption $\tau_{abs}$ and scattering $\tau_{sca}$ optical depths, again using
\cite{Draine1984} and \cite{Laor1993} for the values of $Q_V$,
and then use the approximation that the effective total
absorption optical depth is $\tau_{eff} = (\tau_{abs}(\tau_{abs}+\tau_{sca}))^{1/2}$.

In practice, as soon as dust formation is permitted, the transient
is blacked out, as shown in Figure~\ref{fig:tau}.  This is simply
a consequence of the relatively small $R_f$ and the enormous
amount of mass -- in the first 10 days of this period, $0.1 M_\odot$ of 
material moves beyond the formation radius of order $R_f \simeq 5 \times 10^{15}$~cm.  
This material will have an optical depth of $\tau \simeq 2.5 \kappa$, where 
$\kappa$ is the opacity.  Typical dusty materials have $\kappa_V\sim 10^2$,
which means that the optical depth will be large even if dust formation
is very inefficient.  As the source vanishes in the optical, it would be
replaced by a near/mid-IR source with an initial temperature near the
dust formation temperature.  It would rapidly become colder and fainter
due to the combined effects of the collapsing luminosity and the 
increasing optical depth.  This phase might be observable for a brief
period (weeks) with JWST. 

\section{Discussion}

Despite having very favorable conditions for dust formation, the optical
signature of a failed SNe from a red supergiant will be little 
affected by dust formation.  Failed SNe are, however, very efficient
at producing dust compared to successful SNe.  Keeping all else
fixed, the final grain size in Equation~\ref{eqn:afinal} is
$a_f \propto (v L_*)^{-1}$ because the growth rate at the
dust formation radius scales as $da/dt \propto R_f^{-3} \propto L_*^{-3/2}$ 
while the time scale over which the growth occurs is $R_f/v \simeq L_*^{1/2}/v$.
Since for a failed SNe $v \simeq 10^2$~km/s and $L_* \simeq 10^6 L_\odot$
and for a successful SNe $v \simeq 5000$~km/s and $L_*\simeq 10^9 L_\odot$,
the nominal final size ratio of $a_{fail}/a_{succ} \sim 50000$ gives
a sense of the difference.  The hotter radiation temperature of 
successful SNe further favors the failed SNe.  The one advantage
for the successful SNe is that the heavily metal enriched material
outside the iron core is also ejected, so some fraction of the
ejecta from a SNe has a condensible fraction of $X \sim 10^{-1}$ instead
of $X \sim 10^{-3}$ for the envelope of a red supergiant.   
Somewhat surprisingly, even the very metal rich ejecta
of real SNe have slower dust growth rates than the envelope of a 
failed SN. 

The conditions are so favorable for dust formation that the grains formed
in this scenario are enormous -- up to $a_f \sim 1$~mm in size. 
This is interesting because   
several recent experiments have found evidence for a significant 
population of very large extrasolar dust grains.  Impact detectors
on the {\it Ulysses} and {\it Galileo} spacecraft (\citealt{Landgraf2000},
\citealt{Kruger2007}) measured a significant flux of large grains entering
the heliosphere up to their detection limit of $a \simeq 1\mu$m, and
still larger 10-30$\mu$m grains of apparently extrasolar origin have been 
found by ground-based radar studies of meteoroid re-entry trails 
(e.g. \citealt{Meisel2002}, \citealt{Baggaley2000}). The resulting
rates appear to be consistent with simply extending the roughly
logarithmic distribution of grain masses, $m(dn/dm) \propto m^{-1.1}$,
(e.g. \citealt{Murray2004}, \citealt{Draine2009}) well beyond the standard truncation scale
of the models for dust in the interstellar medium (ISM) at $a \sim 0.1\mu$m.

There has been no natural source of these very large grains because
even the most extreme stellar winds have natural size cutoffs of 
order $1\mu$m.  There there are grains with AGB star compositions 
as large as $10\mu$m found in meteorites, but they are rare 
(\citealt{Zinner1998}).  For still larger grains, ejection from 
proto-planetary disks may be possible (see \citealt{Murray2004}).  
These failed SNe would produce $\sim 10\%$ of all dust mass (assuming 
SN and stars produce comparable amounts of dust) with a mechanism that 
both naturally creates and ejects the grains into the interstellar 
medium (ISM) at relatively low velocities so as to minimize 
sputtering. 

Quantitatively, this would be too little mass to explain the solar
system results.  However, there is almost certainly a problem with
the solar system abundance normalizations because simply extending
the ISM size distribution (e.g. \citealt{Draine2003}) to larger sizes would lead to an abnormal 
average extinction curve ($R_V \simeq 5.8$) and require a greater abundance of condensible 
species (C, O, Fe, etc.) than is available (\citealt{Draine2009}). 
Possible solutions include a local anomaly in the grain size
distribution (\citealt{Draine2009}), propagation effects from the ISM 
into the solar system (\citealt{Slavin2012}) or that the source of the 
grains is not actually extrasolar (\citealt{Belyaev2010}).  
The small dust mass fraction from failed SNe would not create these 
problems.  \cite{Socrates2009} discuss a means of directly
detecting large grains in the ISM as a scattered light halo during 
optical transients.  At present, however, it seems that interest
in these large grains has waned -- a relation to black hole formation,
however bizarre, might provide a motivation to resolve these issues.

\section*{Acknowledgments}
I am grateful to E. Lovegrove and S. Woosley for sharing the results of 
their simulations so that these calculations could be performed, to
N. Murray for discussions of large dust grains, and to T.A.~Thompson
for comments and discussions.

\end{document}